\documentclass[11pt]{article}

\usepackage{amsmath}
\usepackage{amssymb}
\usepackage{caption}
\usepackage{cite}
\usepackage{color}
    \definecolor{darkgreen}{rgb}{0,0.5,0}
    \definecolor{darkblue}{rgb}{0,0,0.6}
    \definecolor{purple}{rgb}{0.4,.2,0.7}
\usepackage{comment}
\usepackage[margin = 2.5cm]{geometry}
\usepackage{graphicx}
\usepackage[hyperfootnotes = false, colorlinks = true, linkcolor = darkblue, citecolor = purple]{hyperref}

\newcommand{\be}{\begin{equation}}
\newcommand{\bea}{\begin{eqnarray}}
\newcommand{\eea}{\end{eqnarray}}
\newcommand{\beq}{\begin{equation}}
\newcommand{\ee}{\end{equation}}
\newcommand{\eeq}{\end{equation}}
\def\la{\label}
\def\ba{\begin{eqnarray}}
\def\ea{\end{eqnarray}}
\def\nref#1{(\ref{#1})}

\begin{document}

\thispagestyle{empty}
\begin{center}
    ~\vspace{5mm}
    
    {\LARGE \bf {The  Page curve  of Hawking radiation  \\ 
    ~ \\
    from semiclassical geometry}} 
    
    \vspace{0.5in}
    
    {\bf Ahmed Almheiri,$^1$ Raghu Mahajan,$^{1,2}$ Juan Maldacena,$^1$ and Ying Zhao$^1$}

    \vspace{0.5in}

    $^1$Institute for Advanced Study,  Princeton, NJ 08540, USA \vskip1em
    $^2$Jadwin Hall, Princeton University, Princeton, NJ 08540, USA
    
    \vspace{0.5in}
    
    {\tt almheiri@ias.edu, raghu\_m@princeton.edu, malda@ias.edu, zhaoying@ias.edu}
\end{center}

\vspace{0.5in}

\begin{abstract}
We consider a gravity theory coupled to matter, where the matter has a higher-dimensional holographic dual. 
In such a theory, finding quantum extremal surfaces becomes equivalent to finding the RT/HRT surfaces in the higher-dimensional theory.
Using this we compute the entropy of Hawking radiation and argue that it follows the Page curve, as suggested by recent computations of the entropy and entanglement wedges for old black holes.
The higher-dimensional geometry connects the radiation to the black hole interior in the spirit of ER=EPR. 
The black hole interior then becomes part of the entanglement wedge of the radiation.
Inspired by this, we propose a new rule for computing the entropy of quantum systems entangled with gravitational systems which involves searching for ``islands'' in determining the entanglement wedge.
\end{abstract}

\vspace{1in}

\begin{center}
{\it Dedicated to the memory of Steven S. Gubser}
\end{center}

\pagebreak

\setcounter{tocdepth}{3}
{\hypersetup{linkcolor=black}\tableofcontents}

\section{Introduction} 

The central question of the information paradox \cite{Hawking:1974sw, Hawking:1976ra} is whether the process of formation and evaporation of a black hole can be described in a unitary fashion.
In particular, unitarity implies that the von Neumann entropy of the Hawking radiation should initially rise but then fall back down, following the so called  ``Page curve'' \cite{Page:1993wv,Page:2013dx}. 
If we replace a black hole by its dual quantum mechanical description, via AdS/CFT, we know that this happens. 
However, one would like to understand how it happens from the gravity point of view.
In gravitational theories we now have formulas that compute von Neumann entropies: the Ryu-Takayanagi formula and its extentions \cite{Ryu:2006bv,Hubeny:2007xt,Engelhardt:2014gca}.  
Indeed,  interesting recent work \cite{Penington:2019npb,Almheiri:2019psf} addressed the closely related problem of studying the evolution of the von Neumann entropy of an evaporating black hole via the minimal quantum extremal surface (QES) prescription of \cite{Engelhardt:2014gca}.
(A QES is a surface that extremizes the generalized entropy functional.)
This amounted to locating and tracking the minimal QES in the evaporating black hole spacetime as a function of boundary time.

The main result of \cite{Penington:2019npb,Almheiri:2019psf}  was that, for an old black hole, past the Page time, the QES is located just behind the event horizon, which thereby excludes most of the interior from the entanglement wedge of the boundary.  
This is in contrast to the situation at early times, where the minimal QES is the trivial surface, and hence the entanglement wedge for a black hole that forms in a pure state includes all of the interior.
For such a black hole, the von Neumann (or fine-grained) entropy increases due to the early Hawking radiation, giving the initial rise of the Page curve, and then decreases once the entanglement wedge moves out to  the  near horizon region.
The early growth of the  boundary entropy is related to the growth of the entropy  of the quantum fields inside the black hole, which is manifest using the nice slices picture of \cite{Polchinski:1995ta, Mathur:2009hf}.
At the Page time, once this region is removed form the entanglement wedge, the entropy can start decreasing.

If we assume that the combined state of the black hole and Hawking radiation is pure, then a Page curve for one implies a Page curve for the other. 
However, this would amount to assuming away the information paradox, and therefore a more direct argument for the Page curve of the radiation is desirable. 
In \cite{Almheiri:2019psf} the entropy of the Hawking radiation was computed in the semiclassical limit and was found to not have a Page curve, reproducing Hawking's original result of information loss.
(See also \cite{Fiola:1994ir} for similar calculations in the CGHS model.) 
As discussed in \cite{Penington:2019npb} and briefly alluded to in \cite{Almheiri:2019psf, Hayden:2018khn}, there is perhaps a way to argue that the QES of the Hawking radiation coincides with that of the black hole.

In this paper, we argue that this is the case by  considering an evaporating black hole in a gravitational theory with holographic matter. 
Namely, we consider a gravity theory with matter described by a quantum field theory that itself has a higher dimensional gravity dual. 
This allows us to compute the entropy of Hawking radiation holographically.
In this case, the prescription of extremizing the generalized entropy in \cite{Engelhardt:2014gca} is equivalent, at leading order, to the standard RT/HRT prescription \cite{Ryu:2006bv,Hubeny:2007xt} of extremizing the area.

The upshot is that the minimization condition in the RT/HRT formula in the higher-dimensional holographic dual ensures that the minimal surfaces of the evaporating black hole and the Hawking radiation must coincide.
The entropy of the radiation computed in this way follows the Page curve, rising initially at early times and then decreasing after the Page time due to the phase transition between two surfaces.  
This is the same transition discussed in the computations of the von Neumann entropy of the black hole in \cite{Penington:2019npb,Almheiri:2019psf}. 

A crucial point is that the region deep inside the black hole interior is connected to the radiation via the extra dimension. 
Therefore, when we search for the minimal QES, we can end up with an entanglement wedge for the radiation that reaches all the way into the interior of the black hole, which is in fact what happens after the Page time. 
This geometric connection is related to the entanglement between the Hawking radiation and the interior modes of the quantum matter. 
We can view this extra dimension as an example of a geometric connection between the radiation and the black hole interior, a realization of ER=EPR \cite{Maldacena:2013xja, Maldacena:2013t1} idea (see also \cite{Susskind:2012uw, Papadodimas:2012aq}).
Our analysis in this paper is restricted to the case of a two-dimensional theory of gravity coupled to a two-dimensional conformal field theory, since in this case the  three-dimensional  gravity dual is very simple, but we expect that the results should generalize without much change to higher dimensions. 

This holographic example suggests a new rule for computing von Neumann entropies of quantum systems entangled with quantum fields  in a gravitational theory. 
This new rule allows for the inclusion of new ``quantum extremal islands" which are regions in the gravitational theory that contain matter entangled with the external quantum system. 
Including these islands can result in a penalty due to their areas, but they can also give larger ``savings'' by reducing the bulk entropy piece of the generalized entropy. 
See also \cite{Hayden:2018khn,Penington:2019npb}  for motivation of this new rule.

This paper is organized as follows. 
In Section \ref{sec:setup}, we discuss theories of gravity coupled to holographic matter and their bulk interpretation. 
We focus on a two-dimensional gravity theory coupled to conformal matter, which itself has a three-dimensional dual. 
In Section \ref{section:QES}, we discuss quantum extremal surfaces and entanglement wedges for an evaporating two-dimensional black hole. 
We discuss the computation of the entropy of the radiation and the black hole, and explain how the Page curve arises.
Finally, in Section \ref{sec:newrule} we discuss a new rule for computing entropies and entanglement wedges for systems entangled with a gravity theory.  
Conclusions are presented in Section \ref{sec:conclude}.

\section{Two-dimensional gravity with holographic matter}
\label{sec:setup}

 
Consider a general two-dimensional theory of gravity.  
The Einstein-Hilbert term in two dimensions is purely topological, but it contributes a constant term to the total entropy of the system.
Nontrivial gravitational dynamics arise when we add an extra ``dilaton''\footnote{In this context, the dilaton should really be called an ``entropion''.} field $\phi$, and consider the general action   
\be \la{GraTd}
I_{\text{grav}}[g^{(2)}_{ij}, \phi] = 
\int d^2y\sqrt{-g}\, \left( \frac{1}{16 \pi G_N^{(2)}} \, \phi R^{(2)} + U(\phi) \right),
\ee
where  we have absorbed a possible purely Einstein-Hilbert term by a shift of $\phi$.  
Adding matter to this system, which is taken to be a CFT$_2$ with some fields collectively denoted by $\chi$, we consider  the total action  
\be \la{GPlusM}
I[g^{(2)}_{ij}, \phi, \chi] = I_{\text{grav}}[g^{(2)}_{ij}, \phi] + I_{\text{CFT}}[g^{(2)}_{ij}, \chi] \, .
\ee
We take this  CFT$_2$ to have a three-dimensional holographic dual.
To justify working in the semiclassical limit in the 2d theory, and to ensure that we have a large-radius dual in 3d, we require that the central charge of the CFT$_2$ satisfies $1 \ll c \ll \frac{\phi}{4G_N^{(2)}}$.\footnote{In certain theories corresponding to near-extremal black holes, we also require that $c\ll \frac{\phi-\phi_0}{4G_N^{(2)}}$, where the $\phi_0$ piece corresponds to the extremal entropy.}
In addition, to have an Einstein gravity dual we need that the CFT is suitably strongly coupled.

\begin{figure}[ht]
\begin{center}
\includegraphics[scale=.45]{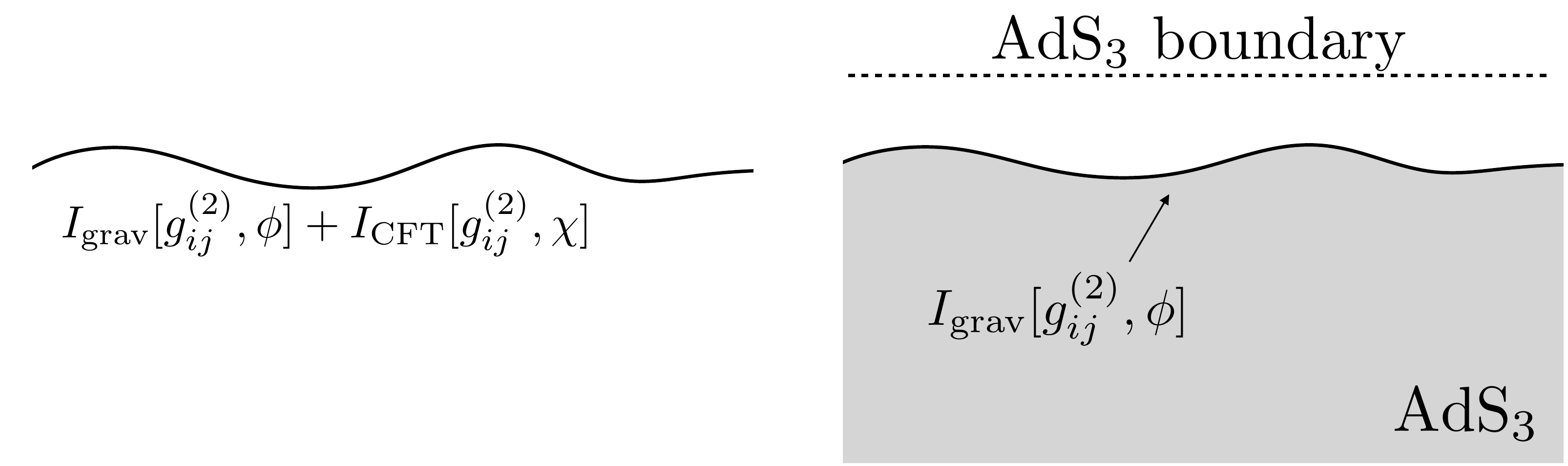}
\end{center}
\caption{On the left, we have a 2d dilaton-gravity theory coupled to a matter CFT$_2$.
The fields of the matter CFT$_2$ are denoted collectively by $\chi$, and this CFT$_2$ is assumed to be holographic.
On the right, we display a 3d geometry obtained by replacing the matter CFT$_2$ with its 3d dual.
This is a version of the Randall-Sundrum setup \cite{Randall:1999vf, Gubser:1999vj}.
On the 2d boundary of this 3d geometry, we have the dilaton-gravity action.
The boundary fields $\phi$ and $g_{ij}^{(2)}$ are also integrated over in the functional integral.}
\label{GravPlusMatter}
\end{figure}

First, let us think about this CFT$_2$ on a fixed background metric $g^{(2)}_{ij}$.
Its three-dimensional dual has a two-dimensional boundary, where the metric obeys the boundary condition 
\be \la{MetBC}
\left. g^{(3)}_{ij} \right|_{\rm bdy} = \frac{1}{\epsilon^2} 
\, g_{ij}^{(2)}\, .
\ee
Here $i,j$ are indices along the boundary (see figure \ref{GravPlusMatter}), and $\epsilon$ is a short-distance cutoff.
According to the  usual rules of AdS/CFT  \cite{Maldacena:1997re,Witten:1998qj,Gubser:1998bc}, a 3d theory with the boundary metric fixed to be $g_{ij}^{(2)}$ is dual to a CFT$_2$ described by the action $I_{ \text{CFT}}[\chi; g^{(2)}_{ij}]$.
The extrinsic curvature of the two-dimensional boundary is related to the stress tensor of the CFT  \cite{Balasubramanian:1999re}.

Next, in order to find the three-dimensional dual to the full action \nref{GPlusM}, we start from the geometry of the previous paragraph,
add a scalar field $\phi$ that lives on the 2d boundary with the action \nref{GraTd}, and integrate over $\phi$ and $g^{(2)}_{ij}$. 
The three-dimensional bulk metric is locally AdS$_3$, with a boundary at a finite location where the 2d theory with with action \nref{GraTd} lives.
We emphasize that unlike usual AdS/CFT, $g^{(2)}_{i j}$ is also integrated over.
This is essentially identical to the Randall-Sundrum model \cite{Randall:1999vf}, and the dynamical boundary brane is called the ``Planck" brane in that paper (for the relationship of the Randall-Sundrum model to holography, see e.g. \cite{Gubser:1999vj}).

The embedding of the Planck brane in AdS$_3$ is determined by using the two-dimensional metric and stress tensor profile from the solution of \eqref{GPlusM}.
Let us describe this in detail. 
Imagine that we have a 2d gravity solution with some profile for the 2d metric and stress tensor
\be \la{Metr}
ds^2 = - e^{ 2 \rho(y) } dy^+ dy^- \, , \quad
T_{y^+ y^+}(y^+) \quad 
\text{and} \quad T_{y^- y^-}(y^-) \, .
\ee 
Here the stress tensor is measured in the flat metric $ds^2 = - dy^+ dy^-$.
The full stress tensor in the original metric \nref{Metr} also contains terms coming from derivatives of $\rho(y)$ which can easily be obtained from the conformal anomaly.
 
It is useful to introduce the coordinate transformations $w^+(y^+)$ and $w^-(y^-)$ that make the stress tensor vanish locally.
These are obtained by solving the equations    
\be
 T_{y^+ y^+} = - { \frac{c}{24\pi} } \{ w^+ , y^+ \} \, ,
 \qquad T_{y^- y^-} = - { \frac{c}{24\pi} } \{ w^- , y^- \}
\, ,
\la{Coorwy}
\ee
where $\{f(y),y\} = \frac{f'''}{f'} - \frac{3}{2} \left(\frac{f''}{f'}\right)^2$ is the usual Schwarzian derivative.\footnote{Note that our convention for the transformation law of the holomorphic stress tensor is $w'(y)^2\, T(w) = T(y) + \frac{c}{24\pi}\{w,y\}$, which differs from \cite{Polchinski:1998rq} by the sign of the Schwarzian term and the normalization of the stress tensor. 
The relation between holomorphic Euclidean coordinates and Lorentzian lightcone coordinates is $y^+ \to \overline{y}$ and $y^- \to -y$.}
The $w^\pm$ coordinates have the property that the stress tensor vanishes in the corresponding flat metric, obtained after a Weyl transformation from (\ref{Metr}):
\begin{align}
ds^2 = - dw^+ dw^-\, , \qquad T_{w^+ w^+} = T_{w^- w^-} = 0. \label{vM}
\end{align}
We therefore observe that the solution \eqref{Metr} is related to the vacuum solution on flat space \eqref{vM} by a combination of Weyl and coordinate transformations.

This determines the location of the Planck brane in the $w^\pm$ coordinates in following way \cite{Banados:1998gg}. 
The vacuum of the holographic CFT$_2$ has pure AdS$_3$ as its associated bulk dual
\be 
\la{vacuum}
ds^2 = \frac{ -dw^+ dw^- + dz_w^2}{z_w^2} \, .
\ee
The stress tensor components $T_{w^+w^+}$ and $T_{w^-w^-}$ vanish for a suface of constant $z_w$.
Therefore, we expect that the geometry near the Planck brane looks similar to (\ref{vacuum}).
The condition we need to impose is that the induced metric on the brane is fixed by (\ref{Metr}) and (\ref{MetBC}), which gives
\begin{align}
- {dw^+ dw^- \over z_w^2}  = - \frac{1}{\epsilon^2} \, e^{2 \rho(y)} dy^+ dy^- \,.
\end{align}
This locates the Planck brane at 
\begin{align}
z_w  = \epsilon \ e^{-\rho(y)} \sqrt{{dw^+ \over dy^+} {dw^- \over dy^-}}\, .
\label{PlanckBr}
\end{align}
After obtaining this we can check that the usual formula for the stress tensor in terms of the extrinsic curvature \cite{Balasubramanian:1999re} gives the one we started in \nref{Metr}. 
These formulas show that once we know the two-dimensional dynamics, given by \nref{Metr}, we can easily find the embedding of the 2d geometry into the 3d one.

The computation of the RT/HRT surfaces is particularly simple in the $(w^+,w^-,z_w)$ coordinates.
We should emphasize that more details about the state of the conformal field theory are encoded deeper into the three-dimensional geometry.
In other words, the RT/HRT surfaces used to compute the various entropies live in a geometry that is not necessarily the same as \nref{vacuum} deeper in the interior, and their areas can depend on the detailed geometry that we  encounter in the interior.

\subsection{Two-dimensional black hole coupled to a bath}

\begin{figure}[t]
    \begin{center}
    \includegraphics[scale=.48]{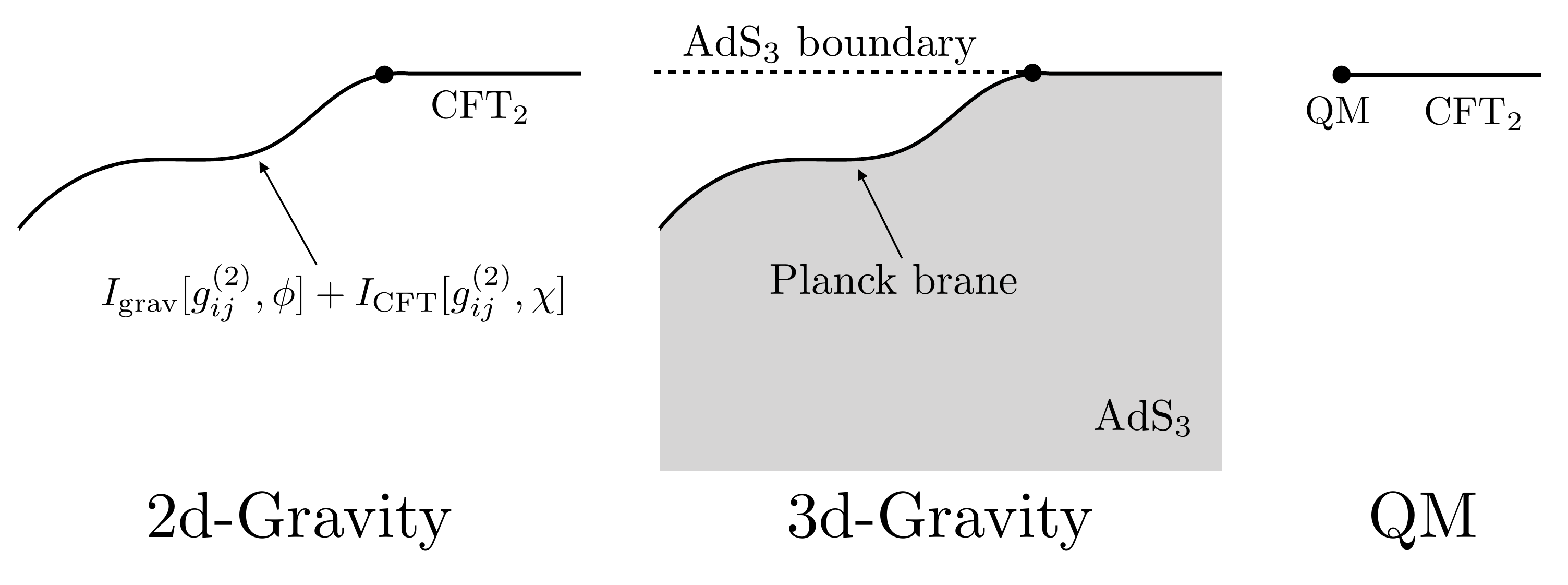}
    \end{center}
    \caption{We sketch three different pictures of the same system.
    The first is a 2d dilaton-gravity theory, plus a matter CFT$_2$, coupled to a bath consisting of the same CFT$_2$. 
    This CFT$_2$ is assumed to have a holographic dual.
    The second is 3d gravity, where we replace the CFT$_2$ by its holographic dual.
    It contains a dynamical boundary metric on the Planck brane.  
    More details about the state of the CFT are encoded deeper inside the 3d geometry.
    The third is the fully quantum mechanical description, where we replace the 2d gravity+matter theory by its quantum mechanical dual. This quantum mechanical system lives at the boundary of the bath CFT.
    In all cases, the thick dot represents the point $\sigma_y=0$.}
    \label{ThreePictures}
\end{figure}

Consider a black hole in the two-dimensional theory \eqref{GPlusM}, which we allow to evaporate into an external `bath'.
For simplicity, we take the bath to be the same CFT$_2$ as the matter sector in (\ref{GPlusM}), but now living on a rigid flat space; see figure \ref{ThreePictures}.
We can think of this setup as a toy model for the case where the dilaton becomes extremely large in some region of the geometry, so that we can neglect backreaction effects and think of the matter as living on a fixed, non-dynamical background. 

We will mostly be interested in the case where the 2d gravitational theory has AdS$_2$ asymptotics.
Prior to coupling in the bath, the matter CFT on this AdS$_2$ spacetime is defined with a conformal boundary condition at the asymptotic AdS$_2$ boundary. 
Prior to the coupling, the bath CFT is also defined on the half line. 
Coupling the two systems amounts to joining them at their boundaries, allowing them to freely exchange stress energy. 
Defining $\sigma_y = (y^+ - y^-)/2$, positive and negative values of $\sigma_y$ correspond to points in the bath and the AdS$_2$ systems, respectively. 

This combined system has three alternative descriptions that are useful, see figure \ref{ThreePictures}.  
\begin{description}
\item[2d-Gravity:]
A two-dimensional gravity-plus-matter theory living on $\sigma_y<0$ coupled to a two-dimensional field theory living on $\sigma_y>0$.

\item[3d-Gravity:] 
A three-dimensional gravity theory in AdS$_3$ with a dynamical boundary (Planck brane) on part of the space ($\sigma_y<0$), and with a rigid boundary on the rest ($\sigma_y>0$). 

\item[QM:]
A two-dimensional CFT on the half-line $\sigma_y>0$ with some non-conformal boundary degrees of freedom at $\sigma_y = 0$.
\end{description}

The first description is the one we have already described in detail.
In the second description, which involves three-dimensional gravity, we replace the CFT$_2$ by its three-dimensional dual. 
This 3d dual has a Planck brane with dynamical gravity on part of the space ($\sigma_y<0$) and the usual UV boundary on the rest of the space ($\sigma_y>0$). 

In the third, fully quantum-mechanical description with no gravity, we replace the 2d black hole by its quantum-mechanical dual (assuming that it has one). 
Then we have a CFT$_2$ living on a half-line coupled to a quantum mechanical system living at $\sigma_y=0$. 
In other words, in the case where the 2d gravity has AdS$_2$ asymptotics, we want to imagine that the entire 2d theory \eqref{GPlusM} arises as the holographic dual of a $(0+1)$-dimensional (nonconformal) quantum-mechanical system. 
After coupling the nearly AdS$_2$ gravity theory to the bath CFT, we get a CFT on the half-line coupled to a holographic quantum-mechanical system on its boundary, as shown in figure \ref{ThreePictures}. 
A higher dimensional version of this set up and its gravity interpretation was discussed in \cite{Karch:2000ct}.

The story so far has been for an evaporating black hole in a general 2d gravity theory with AdS$_2$ asymptotics, coupled to a non-gravitational bath. 
This formalism can be directly applied to the case considered in \cite{Almheiri:2019psf} by specializing to Jackiw-Teitelboim gravity, with the only difference being that we consider matter composed of a holographic CFT$_2$, rather than general matter.
Using their solution for the dynamics of the 2d model, we can follow the simple steps outlined above to find the embedding of the Planck brane and the bath UV brane in the 3d geometry.  

\subsection{Quantum extremal surfaces become ordinary RT/HRT surfaces}

\begin{figure}[t]
    \begin{center}
    \includegraphics[scale=.6]{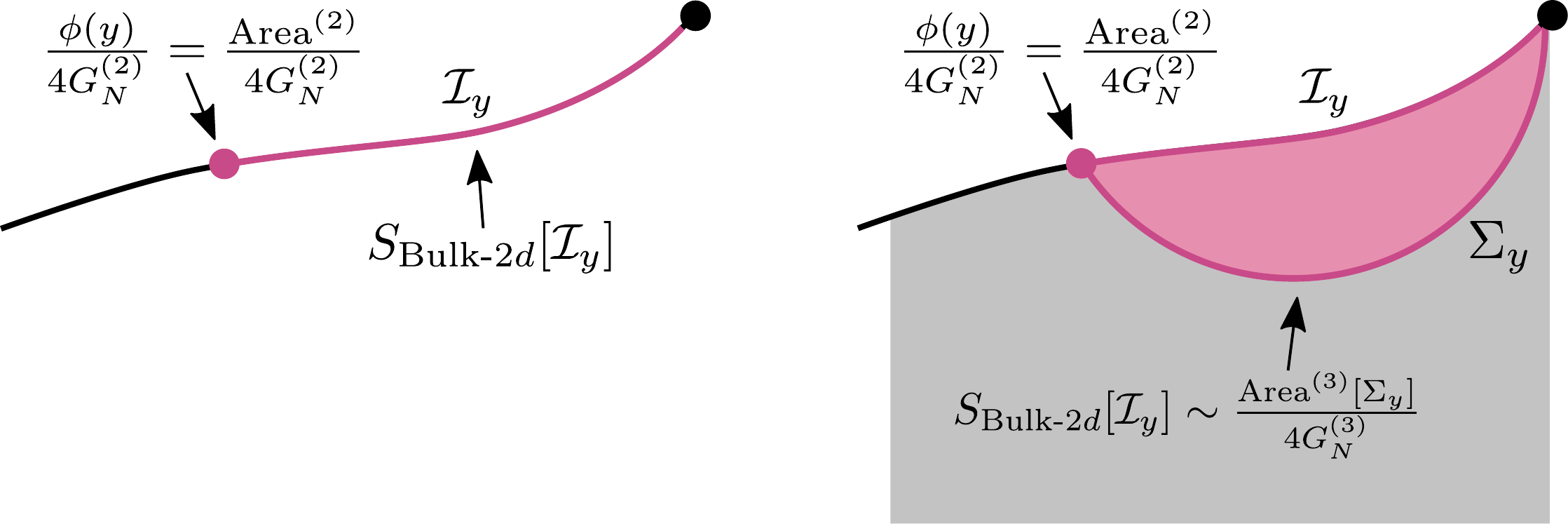}
 \\ (a) ~~~~~~~~~~~~~~~~~~~~~~~~~~~~~~~~~~~~~~~~~~~~~~~~~~(b) 
    \end{center}
    \caption{(a) We have shown, from the 2d perspective, the two contributions $\frac{\phi(y)}{4G_N^{(2)}}$ and $S_\text{Bulk-2d}[\mathcal{I}_y]$ to $S_\text{gen}(y)$ from equation (\ref{GenEN}). (b)  Since the matter CFT$_2$ is holographic, the quantity $S_{\mathrm{Bulk}\text{-}2d}  [{\cal I}_y]$ can be computed using a 3d RT formula.}
    \label{3dsgen}
\end{figure}

In the two-dimensional gravity theory, we can compute the fine-grained entropy  of its quantum-mechanical boundary theory using the prescription of extremizing the generalized entropy \cite{Engelhardt:2014gca}. 
This involves first constructing a quantity similar to the generalized entropy 
\begin{align} \la{GenEN}
S_\mathrm{gen}(y) =  \frac{\phi(y)}{4G_{N}^{(2)}} + S_{\mathrm{Bulk}\text{-}2d}  [{\cal I}_y] \, .
\end{align} 
Here $y$ is a point in the two-dimensional bulk and ${\cal I}_y$ is an interval from the point $y$ to the boundary of the two-dimensional space (or to  some region far away where the dilaton is very large and the theory is very weakly coupled). 
The quantity $S_{\mathrm{Bulk}\text{-}2d}[\mathcal{I}_y]$
is the bulk von Neumann entropy of this interval. 
This bulk entropy includes the entropy coming from the bulk matter fields $\chi$, and also the entropy due to quantum fluctuations of $\phi$ and $g_{ij}^{(2)}$.
Note also that $\phi(y) = \text{Area}^{(2)}$; in two dimensions, the area of a point is the coefficient of the curvature term in \nref{GraTd}.
Figure \ref{3dsgen}(a) shows a slice of the 2d theory including its boundary dual system and indicates where $S_{\text{gen}}$ is evaluated.

Once we construct $S_\text{gen}(y)$ as in \nref{GenEN}, we are instructed to extremize it over the choice of the point $y$. 
And finally, we take the minimum over all such extrema. 
The point $(y_e^+,y_e^-)$ that results from this is called the quantum extremal ``surface."

The contribution of the CFT$_2$ fields $\chi$ to $S_{\mathrm{Bulk}\text{-}2d}[\mathcal{I}_y]$ dominates over contributions from the quantum fluctuations of $\phi$ and $g_{ij}^{(2)}$, since we are assuming that the CFT$_2$ has a large number of degrees of freedom.
Furthermore, since the CFT$_2$  has a holographic dual, the contribution of the $\chi$ fields to the entropy can be computed to leading order using the RT/HRT formula \cite{Ryu:2006bv,Hubeny:2007xt}. 
This involves  finding a minimal or an extremal surface $\Sigma_y$ in the three-dimensional geometry.

The extremal surface $\Sigma_y $ is here just an interval and is shown in blue in  figure \ref{3dsgen}(b).
We therefore have
\begin{align}
    S_\mathrm{gen}(y)  =  {\phi(y)  \over 4 G_N^{(2)} } + 
    S_{\mathrm{Bulk}\text{-}2d}[\mathcal{I}_y]
    ~ \approx ~    
    {\phi(y)  \over 4 G_N^{(2)} }  + 
    {\mathrm{Area}^{(3)}[ \Sigma_y ] \over 4 G_N^{(3)}} \, .
\end{align} 
We used a $\approx$ sign because we are neglecting the contributions from the quantum fluctuations of $\phi$ and $g_{ij}^{(2)}$, and also dropping the subleading 3d bulk entanglement entropy terms.

The extremization of generalized entropy in 2d is equivalent to the standard RT/HRT area extremization in the 3d with a dynamical brane.  
In other words, we look for an area-extremizing surface in 3d with an endpoint on the Planck brane. 
This ``area'' has a contribution coming from the length of the line $\Sigma_y$ as well as a contribution from the dilaton $\phi$ at the Planck brane, see figure \ref{3dsgen}.
We are instructed to extremize the whole thing, which involves also the position of the point $(y^+, y^-)$ on the Planck brane. 
Of course, this observation lends further support to the notion that quantum extremal surfaces are computing von Neuman entropies \cite{Engelhardt:2014gca}, since in this setup it reduces to the simpler RT/HRT proposal in three  dimensions.  This discussion generalizes naturally to higher dimensions.

\section{Entanglement wedges for evaporating black holes and Hawking radiation}   
\label{section:QES}
  
In this section we review the quantum extremal surfaces found in \cite{Almheiri:2019psf} for black holes in JT gravity that evaporate into a non-gravitational bath, and present their corresponding three-dimensional picture.
Our discussion will be out of time order, we will first discuss the late time picture, past the so called ``Page time'' and then discuss the picture for early times. 
We do this because the late time picture is less dependent on the detailed formation history of the black hole. 
It is also the more surprising one, because the entanglement wedge of the radiation will be found to \emph{contain} the region inside the black hole. 
In two dimensions, this region is manifestly disconnected from the radiation, but we show that it is connected to the radiation through the third dimension.
  
\subsection{Late time entanglement wedges}

\subsubsection{Entanglement wedge of the black hole at late times} 
  
\begin{figure}[ht]
    \begin{center}
    \includegraphics[scale=.5]{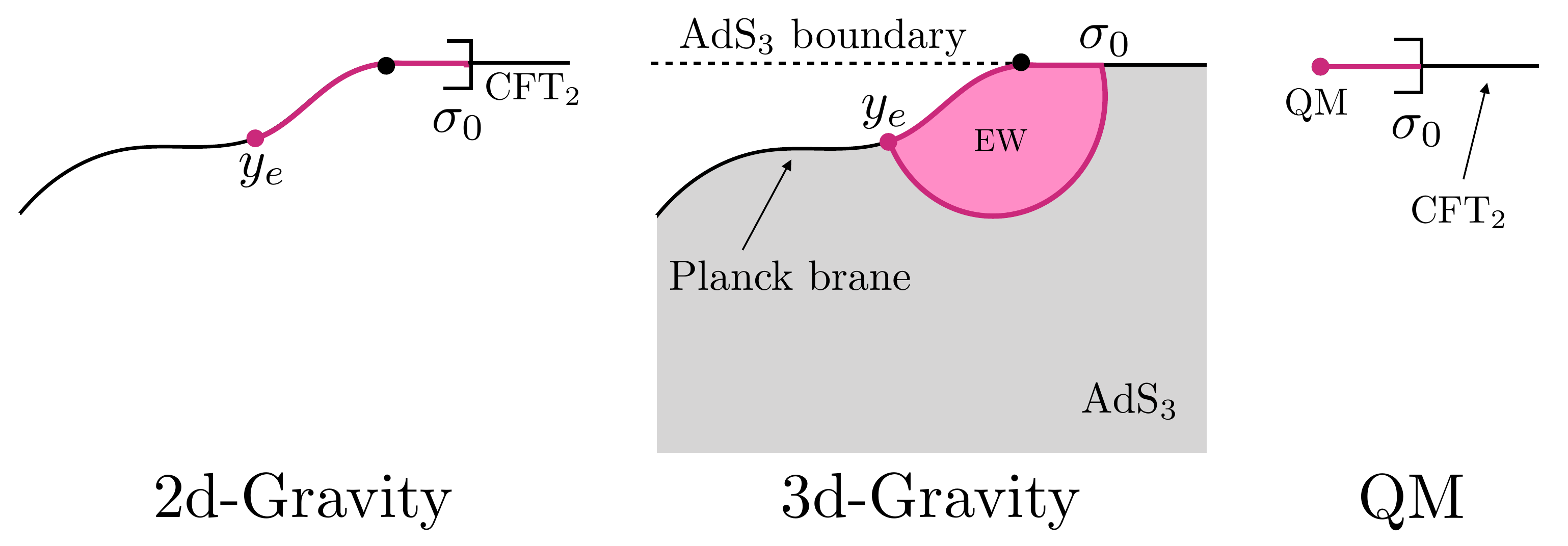}
    \end{center}
    \caption{
     The entanglement wedge for the black hole at late times. 
     We show a spatial slice $\Sigma_\text{Late}$ at some late time that passes through the quantum extremal surface, (see also figure \ref{latetimeslice}).
     In the leftmost picture, we have drawn $\Sigma_\text{Late}$ in the 2d geometry.
     The middle picture is a spatial slice of the three dimensional geometry that ends on $\Sigma_{\rm Late}$ and contains the RT/HRT surface, the pink region being the entanglement wedge. 
     In the rightmost picture, we have an interval that contains the left boundary and whose entropy we are trying to compute.}
    \label{LateEntanglementWedges}
\end{figure}

We start by recalling the position of the quantum extremal surface of the black hole at late times found in  \cite{Penington:2019npb,Almheiri:2019psf}. 
The idea is to consider a subsystem that includes the black hole, or more precisely, the whole gravity region.
From the point of view of the QM description, we imagine that we take a small interval $[0,\sigma_0]$ where $\sigma_0$ is very small, but it includes the quantum-mechanical degrees of freedom at the boundary of the CFT. 
We do this at some late time $t = (y^+ + y^-)/2$.
See figure \ref{LateEntanglementWedges}. 

\begin{figure}[ht]
    \begin{center}
        \includegraphics[scale=.5]{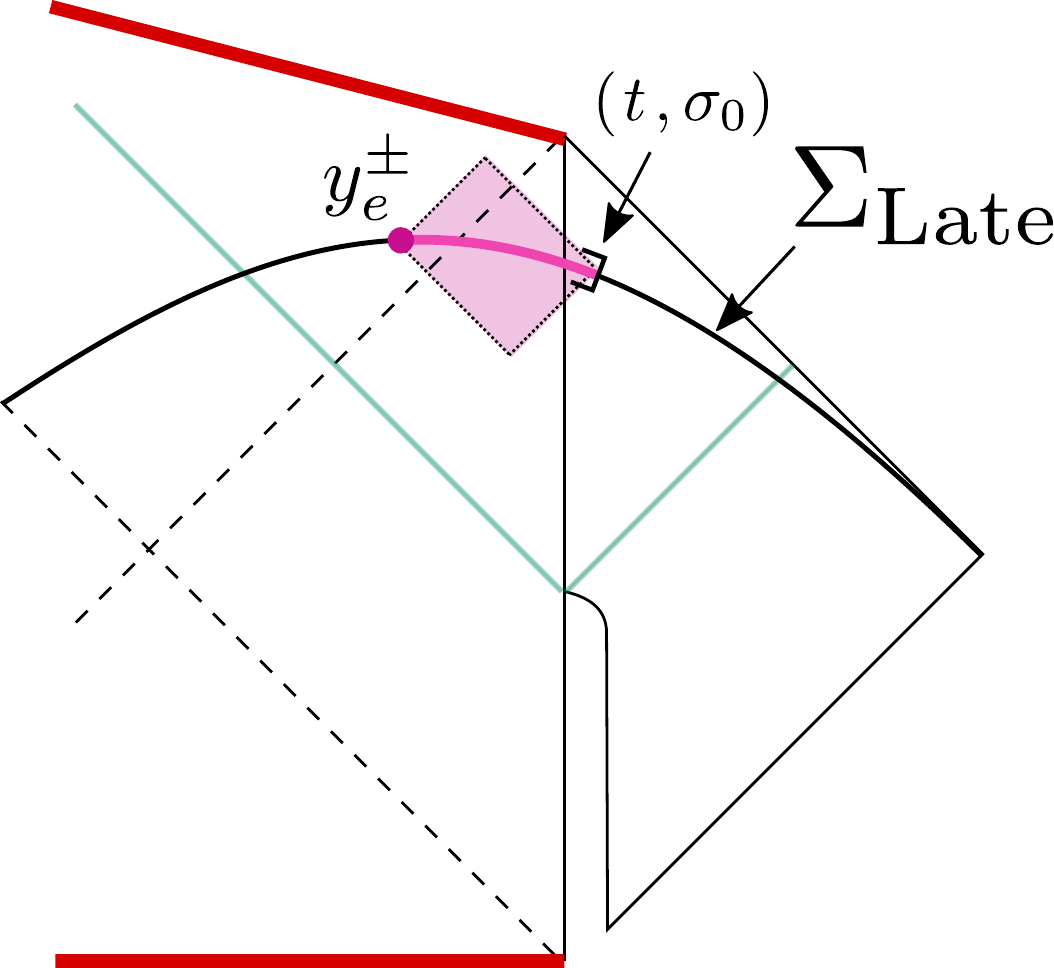}
    \end{center}
    \caption{The spacetime diagram describing the coupling of the black hole to the bath, the energy pulse coming from the moment they are coupled, the formation of the black hole and its subsequent evaporation. We pick some late time nice slice $\Sigma_{\text{Late}} $ and we compute the entanglement wedge for what is to the left of $\sigma_0$. This contains only a portion of the time slice in the interior. We have also displayed the Wheeler de Witt patch, or causal domain of dependence that describes the full entanglement wedge.    }
    \label{latetimeslice}
\end{figure}

The final conclusion of \cite{Penington:2019npb,Almheiri:2019psf} is that the quantum extremal surface is at a point $(y^+_e,y^-_e)$ that lies behind the horizon and is such that a past directed light ray from it would reach the AdS$_2$ boundary at about a scrambling time earlier than the time $t$ at which we are computing the entropy,
\begin{align}
y_e^+ = t - {1 \over 2 \pi T(t)} \log {S_{\text{Bek}}(T(t)) - S_0 \over c} + \ldots\, .
\end{align}
See figure~\ref{latetimeslice}.
Here $T(t)$ is the temperature of the black hole at time $t$, $S_0$ is the extremal entropy (which is assumed to be small), and $S_{\text{Bek}}(T)$ is the Bekenstein-Hawking entropy for a black hole of temperature $T$.
The entanglement wedge is just the causal domain of a spacelike slice going from $(t,\sigma_0)$ to $(y^+_e, y^-_e)$. 
This implies that the computation of the bulk entanglement entropy is just that of an interval.
The answer is slightly nontrivial because the stress tensor is nonzero once we take into account the effects of Hawking radiation,\footnote{In fact, the state at late times is essentially what is called the Unruh vacuum which contains no incoming radiation but contains outgoing radiation.} and hence the map $w^\pm(y^\pm)$ determined from (\ref{Coorwy}) is nontrivial. 
In the $w^\pm$ coordinates, we are just considering an interval in ordinary flat space, and the nontrivial part of the entropy comes from the dependence of the cutoff $z_w$ on the length of the interval \nref{PlanckBr}.

As shown in \cite{Penington:2019npb,Almheiri:2019psf}, 
this leads to an entropy for the black hole of the from 
\be \la{Ent}
   S_{\rm Black~ Hole} (t) = S_{\text{Bek}} (T(t) ) + {\rm logs } \, ,
\ee
where $T(t)$ is the approximate black hole temperature at time $t$.
The ``logs'' denote terms that are logarithmic in the black hole entropy of the initial state.
Note that this entropy is decreasing with time because the temperature is decreasing. 

\subsubsection{Entanglement wedge of the radiation at late times} 
  
\begin{figure}[ht]
    \begin{center}
    \includegraphics[scale=.5]{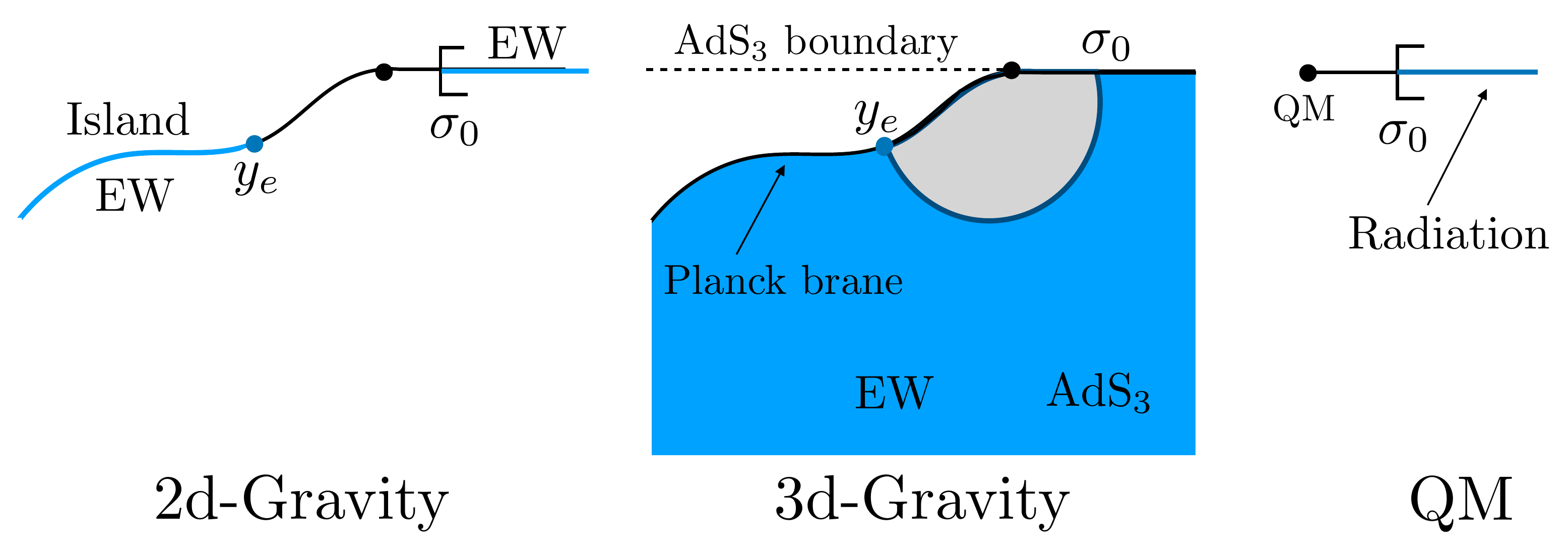}
    \end{center}
    \caption{The analog of figure \ref{LateEntanglementWedges} for the late time entanglement wedge of the \emph{radiation}.
     The rightmost picture depicts the interval whose entropy is being considered.
     The leftmost picture depicts $\Sigma_\text{Late}$ in the 2d gravity picture.
     There is an entanglement island, disconnected from the region where the radiation lives. 
     The middle picture depicts the 3d version, the blue region being the entanglement wedge.}
    \label{RadiationEW}
\end{figure}

Now we consider the entropy of the radiation. 
More precisely, we compute the entropy of the state in the bath CFT outside some point $\sigma_0$, the complement of the interval that we considered for the black hole in the previous subsection, see figure \ref{RadiationEW} right.
\begin{figure}[ht]
    \begin{center}
    \includegraphics[scale=.5]{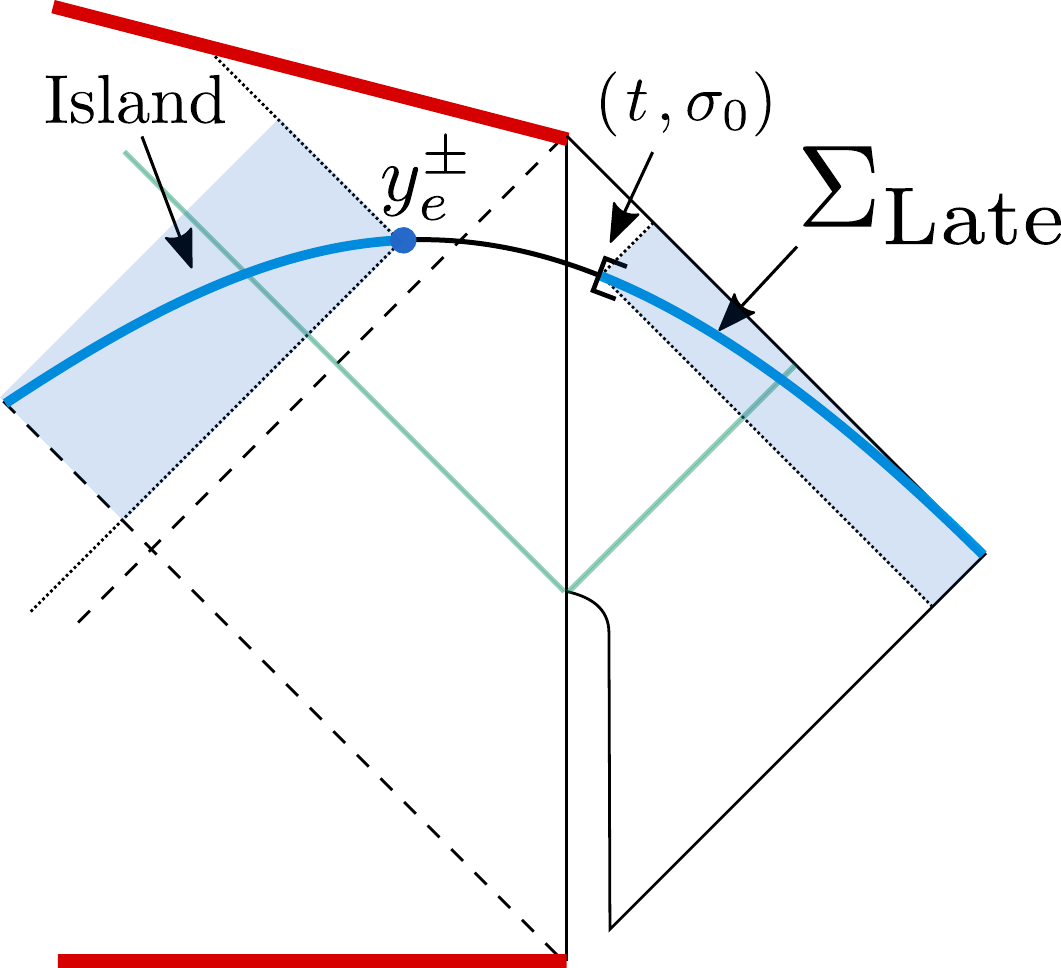}
    \end{center}
    \caption{ Spacetime diagram for the formation and evaporation of the black hole. We consider the entanglement wedges of the radiation. The ``island" is the left wedge, which is disconnected from the right one. }
    \label{latetimeslicerad}
\end{figure}

At first sight this seems straightforward to compute. 
We start from the entire slice and simply trace out everything that is outside the region under consideration. 
This would reproduce the computation in \cite{Hawking:1976ra} which results in an entropy that continues to grow past the Page time.
On the other hand, since the entanglement wedge of the black hole covers only a portion of the interior, it is tempting to think that the rest of the interior should belong to to the entanglement wedge of the radiation \cite{Penington:2019npb, Almheiri:2019psf,Hayden:2018khn}. 
This seems surprising from the two-dimensional point of view where the interior is disconnected from the outside, forming a disconnected  `island', see figure \ref{RadiationEW} left. 
  
When we think about this problem from the three-dimensional point of view we find that the region in the interior is actually connected to the exterior, see figure \ref{RadiationEW} middle. 
If we compute the entropy in the outside CFT by using the standard RT/HRT formula, which instructs us to find the minimal extremal surface, candidate surfaces can end on the Planck brane.
In fact, the extremal surface is essentially the same one as the one we found for the black hole, up to possible IR contributions depending on the precise initial state which not important conceptually, and will be discussed later.
  
This connection through the extra dimensions can be viewed as a realization of the ER=EPR idea that the radiation would be connected to the interior of the black hole.\footnote{In \cite{Maldacena:2013xja} a picture of an ``octopus'' was drawn, that was because the radiation was divided into different parts. Something similar would be the case here too if we were to divide up the radiation in the bath.} 
The extra dimension provides a ``bridge" that connects the ``island" to the ``mainland", the mainland being the CFT interval whose entropy we are computing.
  
The crucial point is that the interior region, which in the two-dimensional picture is disconnected from the outside radiation,  is actually connected through the extra dimension. We expect that this should be a feature of any situation where we have a black hole coupled to holographic matter, even in higher dimensions, $d>2$.\footnote{Higher dimensional black holes in Randall-Sundrum have a variety of phases \cite{Hubeny:2009ru,Figueras:2011gd,Santos:2012he,Santos:2014yja} and only in some cases do they evaporate at leading order.}

This was already suggested as the right prescription on the basis of unitarity in \cite{Hayden:2018khn,Penington:2019npb}. The argument we gave using the holographic example allows us to {\it derive} this fact from the usual rules of RT/HRT surfaces and entanglement wedges. Of course this derivation relies on the assumed correctness of the RT/HRT formula for computing von Neumann entropies. 

\subsection{Early time entanglement wedges}
 
We consider the setup in \cite{Almheiri:2019psf} that starts from a low temperature black hole, initially decoupled from the bath. 
The coupling between the black hole and the bath is turned on at $t=0$.
In order to describe the three-dimensional geometry, we will need some preliminaries.
Many of these points are not essential for the main point of our paper, so the reader who finds them confusing can jump to the next subsection.
   
\subsubsection{The decoupled state}
\la{CFTwB} 
   
\begin{figure}[ht]
    \begin{center}
    \includegraphics[scale=.5]{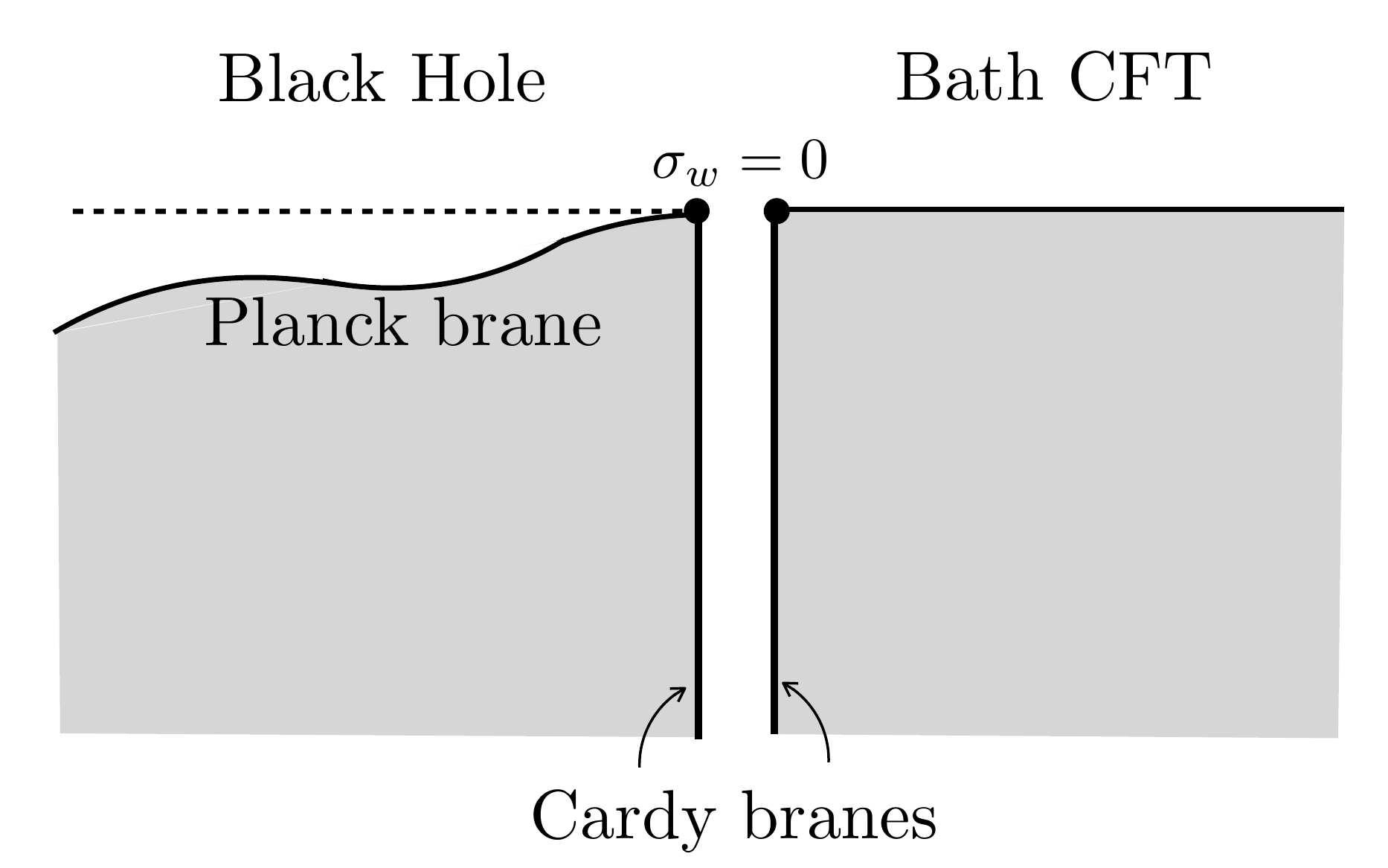}
    \end{center}
    \caption{The decoupled black hole (on the left) and the bath CFT (on the right). We have shown a constant time slice in the 3d geometry. The thick vertical lines are the ``Cardy" branes, which are the holographic dual of conformal boundary conditions.}
    \label{SimpleBrane}
\end{figure}

Suppose that we have a two-dimensional conformal field theory with a simple, Cardy conformal boundary condition \cite{Cardy:1989ir}. 
This type of boundary does not carry any energy, so that $T_{++} = T_{--}$ at the boundary.
The holographic dual of such a boundary condition corresponds to a brane in AdS$_3$ with extrinsic curvature proportional to the metric.
The metric will be \nref{vacuum} with boundary at $\sigma_w = (w^+ -w^-)/2 =0$. 
We will call such a brane a ``Cardy'' brane.
For simplicity, we consider a  brane that goes straight down at $\sigma_w =0$ in the bulk, see figure \ref{SimpleBrane}. 
This brane contains no dilaton or any Ricci curvature term on its surface.\footnote{In principle, the Cardy brane could also go down at an angle, due to a nonzero tension,  and it could also  contain an Einstein term which is topological. Such terms lead to a nonzero  boundary entropy \cite{Affleck:1991tk},  see \cite{Takayanagi:2011zk}.}

Consider an initial state where the bath and black hole are decoupled.
This means that the CFT will have a boundary condition, which is taken to be the conformal boundary condition described above.
The holographic dual consists of two disconnected geometries, see figure \ref{SimpleBrane}. 
    
\subsubsection{Coupling the black hole and the bath CFT} 
 
We now consider coupling the two systems.
(See \cite{Anous:2016kss, Calabrese:2007mtj, Calabrese:2009qy, Asplund:2013zba, Caputa:2019avh, Shimaji:2018czt} for studies of local quenches in two-dimensional CFTs.)
If we were to suddenly couple them at $t=0$, we would get an infinite pulse of energy.
Instead we imagine that we couple them over some time $\Delta t$. 
The state that we get after this will have a pulse of energy of the order $E \propto { c \over \Delta t}$. 
An approximate form for the state can be obtained by considering the Euclidean problem, joining it suddenly and then evolving in Euclidean time for an amount $\Delta t$. 

\begin{figure}[ht]
    \begin{center}
    \includegraphics[scale=.48]{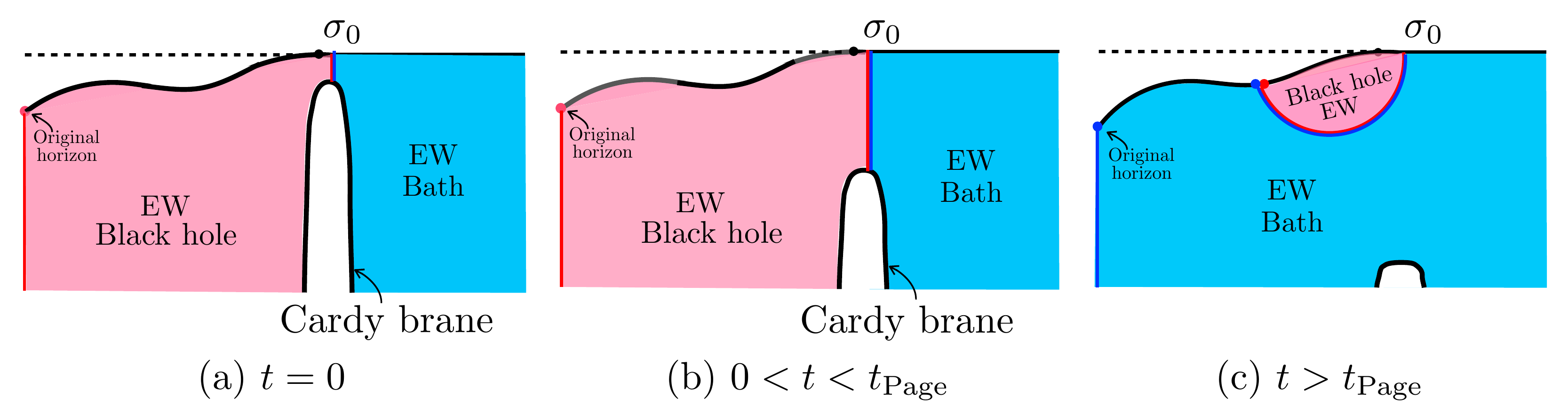}
    \end{center}
    \caption{We show the profile of the simple boundary brane in the bulk at $t=0$. 
    At later times it falls in, in the sense that the physical distance to the boundary increases. 
    We have also shown the black hole entanglement wedges. 
    Where we assumed we started from a black hole at at low temperature where the initial horizon is located where we indicated. The pink region is the entanglement wedge of the black hole, while the complementary blue region is the entanglement wedge of the bath.}
    \label{EarlyEW}
\end{figure}

This leads to a Lorentzian 3d geometry such that, at $t=0$, the Cardy brane is situated somewhere in the bulk, displaced away from the physical boundary towards the AdS$_3$ interior. 
Its precise shape depends on the coordinates used, but a sketch can be seen in figure \ref{EarlyEW}(a). 
As time progresses, this brane falls deeper towards the interior of  AdS$_3$ and becomes more distant from the physical boundary.

We consider an initial black hole which is at a very low (but nonzero) temperature, much lower than the temperature of the black hole that results after the energy pulse falls in.
The entanglement wedge goes very close to the original horizon (the horizon before the energy pulse comes in), and the entanglement wedge of the black hole contains most of the region associated to the black hole, see figure \ref{EarlyEW}(a).

As time progresses, the topology of the figure \ref{EarlyEW}(a) stays similar, but as the brane falls deeper into the bulk, its distance from the boundary increases, leading to a growing entropy, as shown in figure \ref{EarlyEW}(b).
This growing entropy can be physically interpreted as the building up of entanglement between the Hawking modes escaping into the bath and their partners trapped behind the event horizon.
This entropy is equal to the entropy of the radiation and goes as 
\be 
\la{EntropyRa}
    S_{\rm Black ~Hole}(t) \sim S_{\rm Rad}(t) =  
    \frac{\pi c}{6} \int_0^t dt'\,   T(t') = 2 S^i_{\text{Bek}} ( 1 - e^{-{\kappa \over 2}  t })\, ,
\ee
where $T(t')$ is the temperature at time $t'$. 
Here $S^i_{\text{Bek}}$ is the coarse-grained Bekenstein-Hawking entropy of the black hole that forms after the pulse falls in. 
Denoting its temperature by $T_i$, we find that $T(t) \sim  T_i e^{ - {\kappa \over 2} t} $ due to black hole evaporation \cite{Engelsoy:2016xyb}. 
Here $\kappa$ is proportional to $c$ and also to the effective gravitational coupling of the 2d gravity theory, see \cite{Almheiri:2019psf} for details.
       
The important point about \nref{EntropyRa} is that it rises continuously and it saturates at twice the initial black hole entropy.
In the context of figure \nref{EarlyEW} this means that we can neglect the contribution of the leftmost RT/HRT surface that ends at the original horizon. 
The factor of two in \nref{EntropyRa} arises because the Hawking radiation is not adiabatic and it generates coarse grained entropy \cite{Page:2013dx} (see also \cite{Penington:2019npb}).    

\subsubsection{A slightly more precise picture for the entropy of radiation} 
     
Having specified the initial state in more detail, we can be slightly more precise about the entropy of radiation at late times. 
There are a couple of new contributions.

\begin{figure}[ht]
    \begin{center}
    \includegraphics[scale=.47]{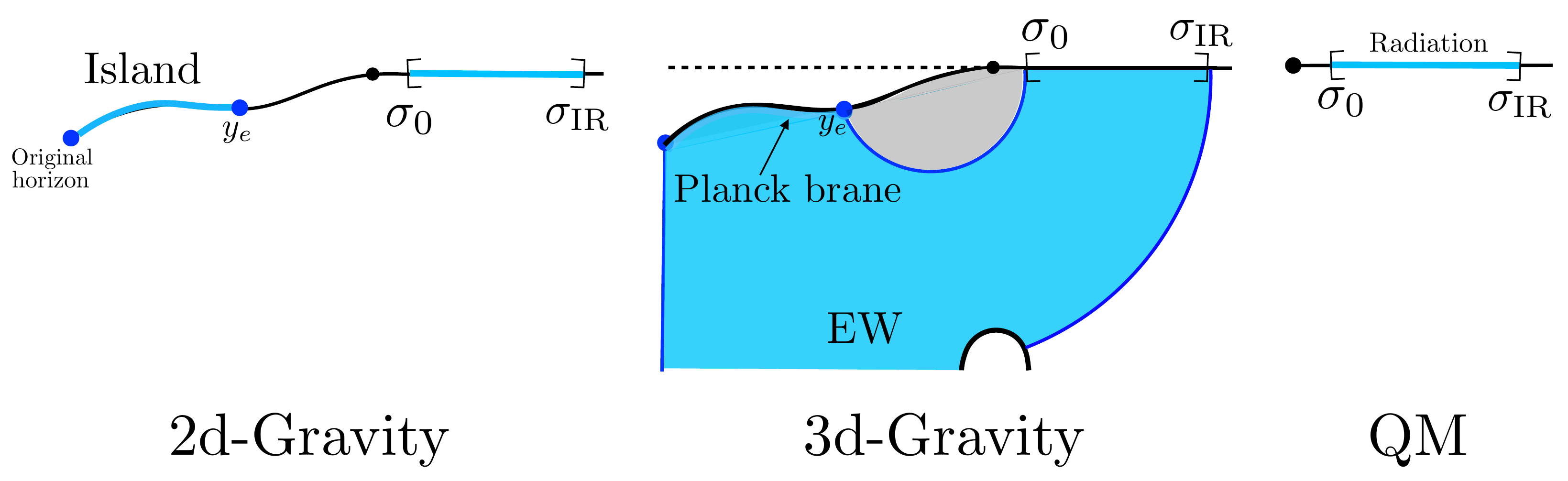}
    \end{center}
    \caption{An improved version of figure \ref{RadiationEW} showing the entanglement wedge of the radiation including a contribution from the initial black hole horizon, and a contribution from the IR cutoff in the CFT. 
    Both these contributions are smaller than $S^i_\text{Bek}$.}
    \label{RaditionEWBetter}
\end{figure}

First, the initial state had some entropy, $S_0$, associated to the horizon of the original low-temperature black hole. 
We are imagining that $S_0$ is much smaller than the Bekenstein-Hawking entropy $S^i_{\text{Bek}}$ of the black hole created by the pulse of energy.
Nevertheless, this implies that, to the left of the ``island", there is also a piece of the RT/HRT surface associated with this original horizon, see figure \ref{RaditionEWBetter} middle.
   
In addition, we can consider radiation as living in a finite interval $[\sigma_0 ,\sigma_{\text{IR}}]$ where $\sigma_{\text{IR}}$ is big enough to contain the entire Hawking radiation, in particular, larger than the total evaporation time.
Then there is a second surface that ends on this endpoint and goes to the ``Cardy" brane.
See figure  \ref{RaditionEWBetter}.
This surface gives rise to a relatively small entropy proportional to ${c \over 6 } \log \sigma_{\text{IR}} \ll S^i_{\rm Bek}$.

\subsection{The Page curve}
\begin{figure}[ht]
    \begin{center}
    \includegraphics[scale=.7]{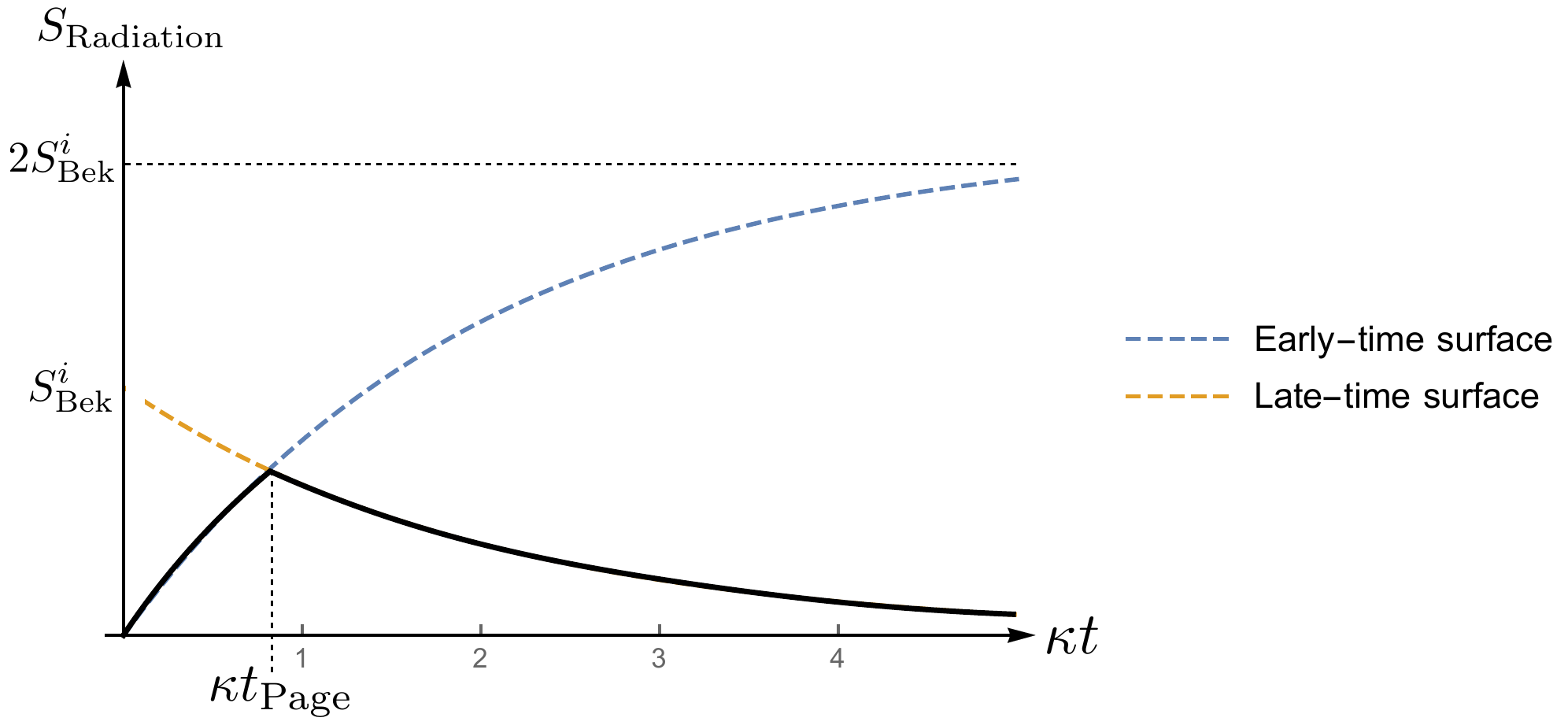}
    \end{center}
    \caption{Sketch of the entropy of the radiation, computed using the early time surface and the late time one. 
    We are assuming that $S_0 \ll S^i_\text{Bek}$. 
    There is a transition between the two at the Page time.
    Here $\kappa$ is proportional to $c$ and also to the effective gravitational coupling of the 2d gravity theory.}
    \label{BHEnt}
\end{figure}

The Page time \cite{Page:1993wv} is defined to be the time where the increasing early-time form of the entropy \nref{EntropyRa} is equal to the decreasing late-time one \nref{Ent}, and the entanglement wedge undergoes a transition.
The minimality condition in the RT/HRT prescription leads to the entropy of the radiation reaching a maximum and then decreasing, see figure \ref{BHEnt}. 
(Recall that we are assuming that $S_0 \ll S^i_{\rm Bek}$.)
Both surfaces exist on either side of the transition, it is just that they may not be the minimal ones.

Note that in the case that the extremal entropy $S_0$ is not negligible, then the minimization process could be different. 
For example, if we had $S_0 \gg S^i_{\rm Bek}$, then the entropy of the radiation will take the short time form for all times.
The reason is that we need to pay a $2S_0$ price to create the ``island'' (becasue of the two endpoints of the island), and this is not favorable when $S_0$ is large.
However, if we keep feeding the black hole with pure matter while the black hole evaporates, eventually the radiation will become entropic enough that it will be worth paying the $2 S_0$ price for the island.

\section{A new rule for computing the entropy when gravitational systems are involved} 
\label{sec:newrule}
 
The calculation of the Page curve using the three-dimensional geometry that we discussed above looks fairly natural from the three-dimensional point of view. 
It does not look very different from other examples where the entanglement wedge extends beyond the causal wedge. 

However it looks surprising from the purely two-dimensional point of view.
There, we have a hybrid system consisting of a black hole coupled to a CFT on a half line.
If we want to compute the entropy of a region in the CFT, then we are strongly tempted to do what Hawking did \cite{Hawking:1976ra}
and restrict attention to the region of the CFT, without including anything else.
However, the holographic computation suggests that we should also include the interior region. 

This suggests that we should introduce a new rule when we compute entropies in effective theories involving gravity. 
If we consider a state in a quantum field theory that is entangled with some other system that lives in a gravity theory, then we should use the RT/HRT/EW method and  include all possible ``islands" that could extremize the entropy.
When we include such an island we will often have to pay a price due to the boundary area of the island, the area term in the gravity theory.
However, there can be situations where the quantum system is entangled with fields inside a closed universe, or the interior of a black hole that has evaporated completely.
In such cases we do not have to pay an area price because we could, in principle, take the whole space. 

\begin{figure}[ht]
    \begin{center}
    \includegraphics[scale=.17]{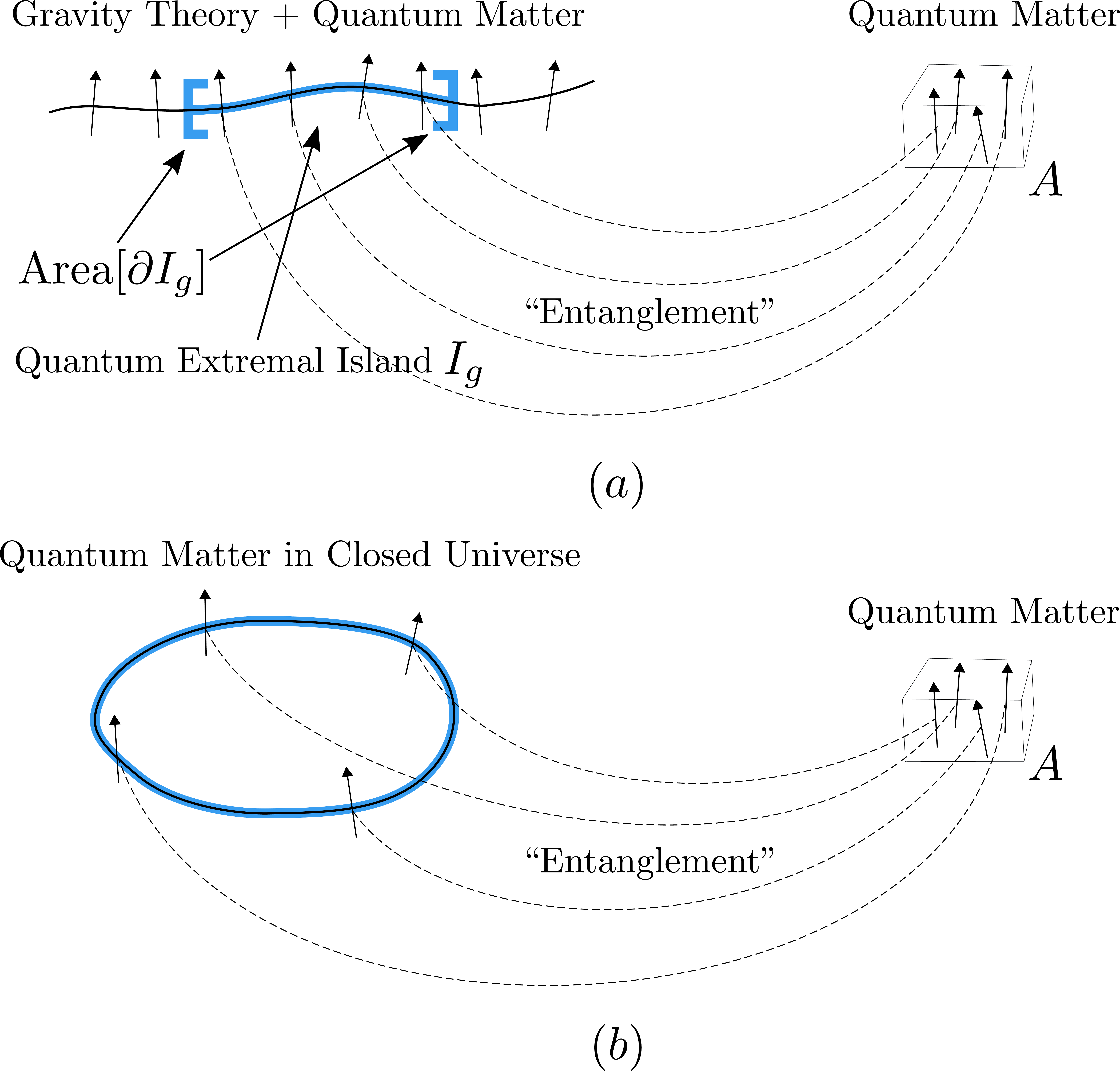}
    \end{center}
    \caption{On the left side we have a quantum system, $A$, entangled with quantum fields living in a dynamical geometry, which we take to have one spatial dimension.  The rule is that we can consider islands, labelled here as blue regions ${\cal I}_g$. In (a) the region ${\cal I}_g$ is a portion of the whole geometry which has a boundary with area $\text{Area}(\partial {\cal I}_g)$. In (b) region ${\cal I}_g$ is the whole universe and it has zero area.}
    \label{islands}
\end{figure}

The prescription is that the actual entropy of some region $A$ in a quantum field theory is given by extremizing a generalized entropy-like functional over islands ${\cal I}_g$ followed by a minimization over all extrema:
\be 
    \la{NewRule}
    S(A) = \underset{{\cal I}_g}{\rm Min}  \, \underset{{\cal I}_g}{\rm Ext}  \left[ S_{\rm eff}(A \cup {{\cal I}_g}) + { \text{Area}[{\partial {\cal I}_g}] \over 4 G_N} \right] \, ,
\ee
where $\text{Area}[{\partial {\cal I}_g}]$ is the area of boundary of the island. The subscript in $S_\text{eff}$ reminds us that we are computing the entropy of the state in semi-classical gravity.
We call the islands ${\cal I}_g$ that extremize this functional \emph{quantum extremal islands}. 
The area contribution form the boundary of the island can be zero if it includes a whole closed universe. 
We imagine minimizing over all possible regions ${{\cal I}_g}$ that could reduce the bulk term for the entropy and also include the areas of such regions, see figure \ref{islands}. 
The subscript in ${{\cal I}_g}$ reminds us that this is an island in a gravity theory.

We should emphasize that this is a rule in the effective field theory. 
If we have the complete and exact state in the quantum system $A$, then we simply use the usual formula for computing the entropy, namely $S(A) = - \text{Tr}(\rho_A \log \rho_A)$, where $\rho_A$ is the density matrix of $A$ in the exact quantum theory.
On the other hand, in the new rule \nref{NewRule} we have a state $\rho^{\rm eff}_{A \cup {{\cal I}_g}}$ which is a semiclassical gravity state.

Note,  in  particular, that we cannot ``forget'' how we obtained the state. 
When we obtained it using a gravity theory, we should also compute its entropy using the gravity formula \nref{NewRule}. 
The system $A$ could be a simple spin chain. 
But if we obtained the state in this spin chain by collecting Hawking radiation from a black hole, we should still include the interior in computing the entropy. 
If we solved the black hole evolution exactly and we give the exact state in the spin chain, then we can use the standard von Neumann formula for the entropy of this spin chain. 
We can also view the geometry as part of the specification of the state and therefore we need to use the proper RT/HRT/EW prescription to compute its entropy. 

Another perspective is the following. 
If we have a configuration that has a quantum system in one region of space and has dynamical gravity in some other region of space, even to compute the entropy of the quantum system, we should still use the rules for gravity theories.
In other words, if we transfer the Hawking radiation to a set of spins, using the gravity theory, then the combined configuration of spins and black hole spacetime is the full state. When we compute the density matrix we need to follow the rules of gravity to find the entanglement wedge of the spins and its corresponding entropy.
The full configuration can be thought of as a tensor network preparing the state of the spins \cite{Swingle:2009bg}\cite{Hartman:2013qma}. 
So, when we compute the entropy we need to take into account the full network. 
More precisely, the entanglement displayed in figure \ref{islands} should be viewed as internal links in that tensor network.

If the whole computation occurs in a holographic theory, such as in the usual Engenhardt-Wall \cite{Engelhardt:2014gca} quantum extremal surface prescription, then islands should also be allowed in the bulk when we perform the search for a minimal extremal surface.

Notice that the size of the island can be very large, even if the entanglement is relatively small. 
So, this new rule says, for example, that if we have a single spin-$1/2$ particle entangled with another spin-$1/2$ particle in a very large closed universe which is otherwise in a pure state (its total bulk entropy arises only due to the spin half particle), then the entanglement wedge of the original spin half particle includes the whole closed universe, see figure \ref{islands}(b).
Now, this is surprising because the whole closed universe could be in different pure states. 
However, if we do not know which pure state we have, and we want to include those alternatives, we generate a mixed state and therefore  the entropy inside this universe increases and we would not get that the entanglement wedge of the outside spin includes the whole universe. 
In other words, this formula is not saying that we can learn about the state of the whole closed universe by just looking at a single spin-1/2 particle. 
Note that by this rule, the entanglement wedge could potentially also include a whole set of additional closed universes in pure states, but which would not contribute to the generalized entropy at all, and we continue to not learn anything about them.

It is possible that there is also an exact description of the gravity theory where \nref{NewRule} is exact. 
(As discussed in \cite{Penington:2019npb}, such a description might involve something similar to a final state projection \cite{Horowitz:2003he}.)
We are calling (\ref{NewRule}) as an effective prescription, because in general, we can only view the gravity theory as an effective field theory. 
It is of course very interesting that gravity still knows about the fine-grained von Neumann entropy of the state. 
We can view this as evidence that there is some bulk theory that contains the precise information about the state. 
  
Notice that this set of ideas is also connected to the Bekenstein area bound in an interesting way. 
The Bekenstein bound says that the entropy of a region bounded by some surface $\mathcal{S}$ should be smaller than the area of $\mathcal{S}$.
This bound is clearly violated if we consider $\mathcal{S}$ to be the horizon of an old evaporating black hole and use semiclassical reasoning.
However, in such situations when the bound is violated, we expect that there is a nontrivial quantum extremal surface $\mathcal{E}$ inside $\mathcal{S}$, and now the $S_{bulk}$ term should only contain the entropy between $\mathcal{E}$ and $\mathcal{S}$, instead of the entropy of the entire region bounded by $\mathcal{S}$.
The entanglement wedge thus might be smaller than the entire region inside $\mathcal{S}$, and the large entropy remains outside the entanglement wedge. 
This suggests a modified version of the Bekenstein bound stating that:

\begin{flushleft}
{\it The generalized entropy of the entanglement wedge inside a region with boundary ${\cal S}$   should be less than ${\rm Area}({\cal S}) /(4 G_N)$.}  
\end{flushleft}

Let us discuss this in more detail. 
This statement presumes that we can consider an arbitrary surface $\mathcal{S}$ and view it as being a holographic-type boundary where a quantum system lives.
Then we consider a candidate quantum extremal surface $\mathcal{E}$ inside the region bounded by $\mathcal{S}$. 
This modified bound would then follow simply from the minimization prescription, if we extremize over the choice of $\mathcal{E}$.
This is due to the fact that a very ``thin'' entanglement wedge, where $\mathcal{E}$ almost coincides with $\mathcal{S}$, would be an example of the surfaces over which we are extremizing. 
Since it is very thin, it does not capture any bulk entropy, and thus the generalized entropy functional evaluates to $\text{Area}(\mathcal{S})/(4G_N)$ on this very thin entanglement wedge.
Therefore, the true minimal extremal surface should have smaller generalized entropy. 
 
We should emphasize how surprising the RT/HRT/EW formulas are \cite{Ryu:2006bv, Hubeny:2007xt, Engelhardt:2014gca}. 
They claim that we can compute the fine-grained entropy just by looking at the effective gravity theory.
This is surprising because one might expect that we need some knowledge that goes beyond the effective field theory (the details of the UV completion of the theory, for example) in order to compute the fine-grained entropy.
It is expected that if one wanted to compute detailed matrix elements of the density matrix, then we would need to have an accuracy of order $e^{-S}$, where $S$ is the black hole entropy. 
Such computations are expected to be very sensitive to the microscopic details of the theory. They are also  likely to involve other topologies, as in the long time discussion in \cite{Saad:2018bqo,Saad:2019lba}.
However just the effective field theory is smart enough to know about the correct entropy of Hawking radiation. 
One simply has to apply the correct prescription for its entropy!\footnote{Progress on a direct derivation of \eqref{NewRule} has been reported in \cite{PeningtonWIP,AlmheiriWIP}. }

\section{Conclusions}
\label{sec:conclude}

To summarize,
we have considered the computation of the entropy of Hawking radiation for an evaporating black hole. 
We studied a two-dimensional black hole coupled to a two-dimensional matter CFT, where this matter CFT has a holographic dual. 
In this case, the quantum extremal surface prescription of \cite{Engelhardt:2014gca} reduces to the usual RT/HRT \cite{Ryu:2006bv,Hubeny:2007xt} prescription in the bulk geometry.
When interpreted in terms of the two-dimensional theory, the entanglement wedge has an ``island'' in the black hole interior. 
The appearance of this island was previously noted in the computation of the entanglement wedge of the black hole in \cite{Penington:2019npb,Almheiri:2019psf}. 
This island is connected to the exterior, where radiation lives, by the extra dimension of the holographic theory. 
One can view the resulting geometry as a particular realization of the ER=EPR \cite{Maldacena:2013xja, Maldacena:2013t1} idea (see also \cite{Susskind:2012uw, Papadodimas:2012aq}). 

The radiation is described by a density matrix living in an ordinary quantum field theory without gravity. 
So in principle, we can compute its entropy using the standard formula $S = - \text{Tr}(\rho \log \rho)$. 
This presumes that we know the state $\rho$ precisely enough. 
If the final state was obtained via the evaporation of a black hole, or some other process involving gravity, then, when we trace out the rest in the semiclassical approximation, we do not get a precise enough approximation for $\rho$. 
In fact, we would get the standard Hawking answer \cite{Hawking:1976ra}. 
However, the full geometry {\it does} contain enough information to compute the fine-grained entropy, but perhaps not enough to compute the precise individual matrix elements of the density matrix. 
For this, one needs to make up a new rule for how to compute entropies for quantum systems that are entangled with gravitational systems. 
We need to consider the addition of ``islands'' and use the quantum extremal surface prescription  \cite{Ryu:2006bv,Hubeny:2007xt,Engelhardt:2014gca} to find the precise shape for the island. 
 
The case with holographic matter that we have considered here makes this new rule plausible. 
But we expect it to hold even in cases where matter is given by free fields which do not have a standard Einstein gravity dual. 

Given the existence of these islands, it would be interesting to see whether there exists a method   to extract the information contained within them that has a clear bulk interpretation, in the spirit of what was discussed for the Hayden-Preskill problem \cite{Hayden:2007cs} in \cite{Gao:2016bin,Maldacena:2017axo}.

Notice that these  islands, together with a suitable statement about entanglement wedge reconstruction, suggests a degree of non-locality for the theory. So it would be interesting to understand better how it is compatible with ordinary local gravitational bulk physics.

Note that for the CGHS model 
\cite{Callan:1992rs,Russo:1992ht,Russo:1992ax,Strominger:1994tn, Fiola:1994ir}, which is asymptotically flat,  we expect a very similar story. We need to compute the entropy of radiation in the asymptotic region, where the dilaton is very large. Again we expect the development of an island in the black hole interior. In the case that the matter is a CFT with a holographic dual, the island is connected to the radiation through the extra dimension. 

Finally, even though we restricted our analysis to the simple case of two dimensions, we expect that the results should be similar in higher dimensions when we have a gravity theory in $d$ dimensions which contains matter that is holographically dual to a $(d+1)$-dimensional geometry.

\subsection*{Acknowledgments}

We would like to thank Netta Engelhardt, Daniel Harlow, Andreas Karch, Donald Marolf, Henry Maxfield, Geoffrey Penington, Steve Shenker, Douglas Stanford, Sandip Trivedi, and Edward Witten for useful conversations. 
A.A. is supported by funds from the Ministry of Presidential Affairs, UAE.
R.M. is supported by US Department of Energy grant No.\ DE-SC0016244.
The work of R.M. was performed in part at the Aspen Center for Physics, which is supported by National Science Foundation grant PHY-1607611.
J.M. is supported in part by U.S. Department of Energy grant DE-SC0009988 and by the Simons Foundation grant 385600.
Y.Z. is supported by the Simons foundation through the It from Qubit Collaboration.

\bibliographystyle{apsrev4-1long}
\bibliography{main}
\end{document}